# Manifestation of geometric resonance in current dependence of AC susceptibility for unshunted array of Nb-AlO$_x$-Nb Josephson junctions

V. A. G. Rivera[1], S. Sergeenkov[2,*], E. Marega[1] and F.M. Araújo-Moreira[2, #]

[1]*Instituto de Física de São Carlos, USP, Caixa Postal 369, 13560-970, São Carlos, SP, Brasil*

[2]*Grupo de Materiais e Dispositivos, Departamento de Física, UFSCar, Caixa Postal 676, 13565-905, São Carlos, SP, Brasil*

**Abstract**

By improving resolution of home-made mutual-inductance measurements technique, a pronounced resonance -like structure has been observed in the current dependence of AC susceptibility in artificially prepared two-dimensional array of unshunted Nb-AlO$_x$-Nb Josephson junctions (JJA). Using a single-plaquette approximation for JJA model, we were able to successfully fit our data assuming that resonance structure is related to the geometric (inductive) properties of the array.

**Keywords:** Josephson junction arrays; AC susceptibility; inductance related effects

---

[*] Corresponding author: phone: (16) 260-8205; fax: (16) 260-4835; E-mail: sergei@df.ufscar.br
[#] Research Leader; E-mail: faraujo@df.ufscar.br

## 1. Introduction

Probably one of the most promising tendencies in modern development of artificially prepared Josephson Junction Arrays (JJA) is based on investigation of intricate correlations between their transport and magnetic properties which include, in particular, simultaneous measurements of current-voltage characteristics (CVC) and AC susceptibility (for genral reviews on the subject matter, see, e.g., [1-4] and further references therein). Among the numerous spectacular phenomena recently discussed and observed in JJA, we could mention the dynamic reentrance of AC susceptibility [5-11], geometric quantization [12], flux driven temperature oscillations of thermal expansion coefficient [13], current driven giant fractional Shapiro steps [14-16], etc. At the same time, successful adaptation of the so-called two-coil mutual-inductance technique to impedance measurements in JJA provided a high-precision tool for investigation of the numerous magnetoinductance related effects in Josephson networks [17-19].

In this Letter we present experimental evidence for manifestation of novel geometric effects in magnetic response of high-quality ordered array of unshunted Nb-AlO$_x$-Nb junctions under application of AC current. We observed a pronounced resonance -like structure in the current dependence of AC susceptibility which was discussed within the single-plaquette approximation and related to inductance properties of our array.

## 2. Experimental Results

High quality ordered SIS type unshunted array of overdamped Nb-AlO$_x$-Nb junctions has been prepared by using a standard photolithography and sputtering technique [1]. It is formed by loops of niobium islands linked through 100x150 tunnel junctions. The unit cell of the array has square geometry with lattice spacing a=46μm and a single junction area of S=5x5 μm$^2$ (see Fig.1). The critical current for the junctions forming the array is $I_C(T)$=150 μA at T=4.2K. Given the values of the junction quasi-particle resistance R=10Ω, inductance L=μ$_0$a=64pH, and capacitance C=1.2fF, the geometrical and dissipation parameters are estimated to be $\beta_L(T)=2\pi LI_C(T)/\Phi_0$=30 and $\beta_C(T)=2\pi CR^2I_C(T)/\Phi_0$=0.05 at T=4.2K, respectively. The latter estimate suggests that our array can be effectively used, for example, in rapid single flux quantum logic circuits and programmable Josephson voltage standards [20,21].

For comparative study of current driven effects in our array, we measured their transport and magnetic properties. For both purposes, AC current $I_{ac}(t) = I\sin\omega t$ (with amplitude 0<I<5mA and fixed frequency $\omega = 40kHz$) was applied parallel to the plane of the array and normally to the Josephson current $I_J$ (see Fig.2 for the sketch of the setup used in the present experiments). The measurements of CVC V(I) were made using homemade experimental technique with a high-precision nanovoltmeter [21-23]. Some typical results for the obtained V(I) dependencies in our array (taken at two distinctive temperatures and exhibiting a noticeable nonlinear behavior) are shown in Fig. 3.

To measure the current induced response of complex AC susceptibility $\chi(I)$ with high precision, we used a home-made susceptometer based on the so-called screening method in the reflection configuration [5,6,9-12]. We observed a characteristic oscillating dependence of the zero-field (B=0) normalized susceptibility $\chi(I)/\chi(0)$ on applied current I as well as a pronounced resonance-like peak around I=2mA which is clearly seen in Fig.4.

## 3. Discussion

Turning to the interpretation of the obtained experimental results, it is important to mention that magnetic field dependence of the critical current of the array (taken at T=4.2K) on DC magnetic field B (parallel to the plane of the sample) exhibits [5-11] a sharp Fraunhofer-like pattern characteristic of a single-junction response, thus proving a rather strong coherence within array (with negligible distribution of critical currents and sizes of the individual junctions) and hence the high quality of our sample.

Due to the well-defined periodic structure of our array, it is quite reasonable to assume that the experimental results could be understood by analyzing the dynamics of just a single unit cell (plaquette) of the array. As we shall see, theoretical interpretation of the presented here experimental results based on single-loop approximation is in excellent agreement with the observed behavior.

In our analytical calculations, the unit cell is the loop containing four identical Josephson junctions. If we apply a DC magnetic field B and an AC current $I_{ac}(t) = I\sin\omega t$ parallel to the JJA, the total magnetic flux $\Phi(t)$ threading the four-junction superconducting loop is given by $\Phi(t) = BS + LI_{ac}(t)$ where L is the loop (plaquette) inductance, and S=5x5 µm$^2$ is a single junction area (see Fig.1).

To properly treat the current mediated magnetic properties of the system, let us introduce the following model Hamiltonian [4,12]

$$H(t) = J[1 - \cos\varphi(t)] + \frac{\Phi^2(t)}{2L} \qquad (1)$$

which describes the tunneling (first term) and inductive (second term) contributions to the total energy of the four-junction plaquette. Here, $\varphi(t) = \frac{2\pi\Phi(t)}{\Phi_0}$ is the gauge-invariant superconducting phase difference across the ith junction [4-12], and the corresponding Josephson current (shown in Fig.2) is given by $I_J = I_C \sin\varphi$ where $I_C = 2\pi J/\Phi_0$ is the critical current.

The seeking dependence of zero-field (B=0) susceptibility on the amplitude of the applied current I can be defined as a time average over the period $\tau = 2\pi/\omega$ (Cf.[4]):

$$\chi(I) = \frac{1}{\tau}\int_0^\tau dt \cos\omega t \left[-\frac{\partial^2 H(t)}{\partial B^2}\right]_{B=0} \qquad (2)$$

Solid line in Fig.4 shows the best fit of the measured normalized susceptibility $\chi(I)/\chi(0)$ using Eq.(2) with $\chi(0)$=−0.72SI, L=64pH, and ω=40kHz. A remarkable agreement between the experimental points and theoretical curve justifies *a posteriori* the adopted here single plaquette approximation scenario. Furthermore, a careful analysis of Eq.(2) reveals that the observed resonance occurs for $\omega \cong \omega_r(I)$, where ω=40kHz is the frequency of the applied AC current $I_{ac}(t) = I\sin\omega t$ and $\omega_r(I) = \left(\frac{2}{\pi}\right)\frac{\omega L_\omega I}{\Phi_0}$ is the current induced resonance frequency with $L_\omega = \frac{L}{1+\omega^2\tau^2}$ being the dynamical inductance of the plaquette [2,17,18]. More precisely, near the resonance, Eq.(2) can be approximated (with good accuracy) by the Fraunhofer-type dependence $\chi(I) \approx -\chi_0 \frac{\sin(\omega-\omega_r)\tau}{(\omega-\omega_r)\tau}$ with $\chi_0 = \pi S^2 I_C/2\Phi_0$. Recalling that $\omega\tau = 2\pi$ and L=64pH, we find that the resonance condition $\omega \cong \omega_r(I)$ for our array will be satisfied for the amplitude of the AC current equal to $I \cong 2mA$, in excellent agreement with the observations (see Fig.4).

Finally, it is important to emphasize that the conventional AC Josephson effect with frequency $\omega_J(I) = \frac{2eV(I)}{\hbar}$, related to the above-discussed nonlinear CVC law V(I) in our array, is too high to account for the observed resonant structure of AC susceptibility. Indeed, according to Fig.3, the resonant current of I=2mA corresponds to the voltage of V=175mV (see the intersection of dotted lines in Fig.3 for T=4.2K) which gives $\omega_J(I) \approx 10^{14} Hz \gg \omega$ for the estimate of the Josephson frequency in our array.

## 4. Summary

In summary, we reported on observation of pronounced resonance-like structure in current dependence of AC susceptibility for SIS-type ordered array of unshunted Nb-AlO$_x$-Nb Josephson junctions. The origin of the observed phenomenon was discussed within a single-plaquette approximation and attributed to manifestation of geometric (inductance related) properties of the array.

**Acknowledgments**

We are very grateful to R.S. Newrock, P. Barbara and C.J. Lobb for helpful discussions and for providing high quality SIS samples. This work has been financially supported by the Brazilian agencies CNPq, CAPES and FAPESP.

**Figure Captions**

**Fig.1:** SEM photography of a single plaquette (consisting of four junctions) for the studied array of unshunted Nb-AlO$_x$-Nb Josephson junctions.

**Fig.2:** A simplified sketch of the experimental setup showing the directions of applied I$_{ac}$ and Josephson I$_J$ currents in the array.

**Fig.3:** Current-voltage characteristics of the array taken at two distinctive temperatures.

**Fig.4:** The current dependence of the normalized zero-field susceptibility $\chi(I)/\chi(0)$ (taken at T=4.2K), along with the best fit (solid line) according to Eq.(2).

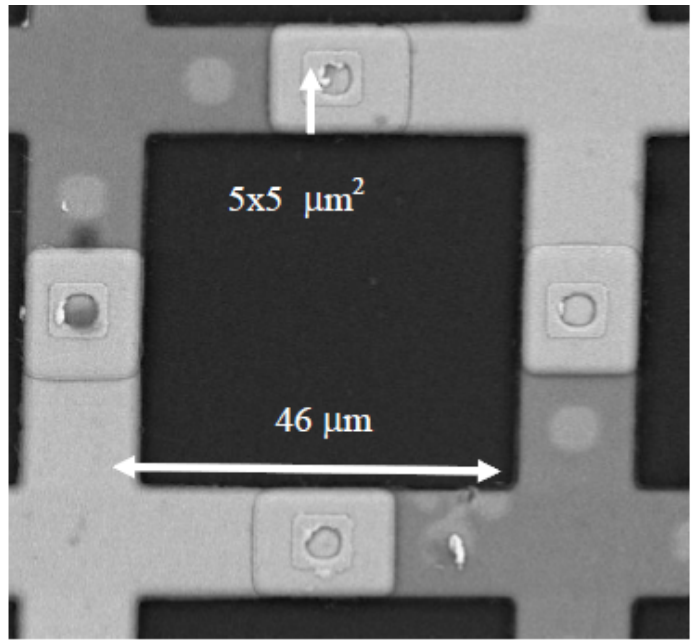

**Fig.1**

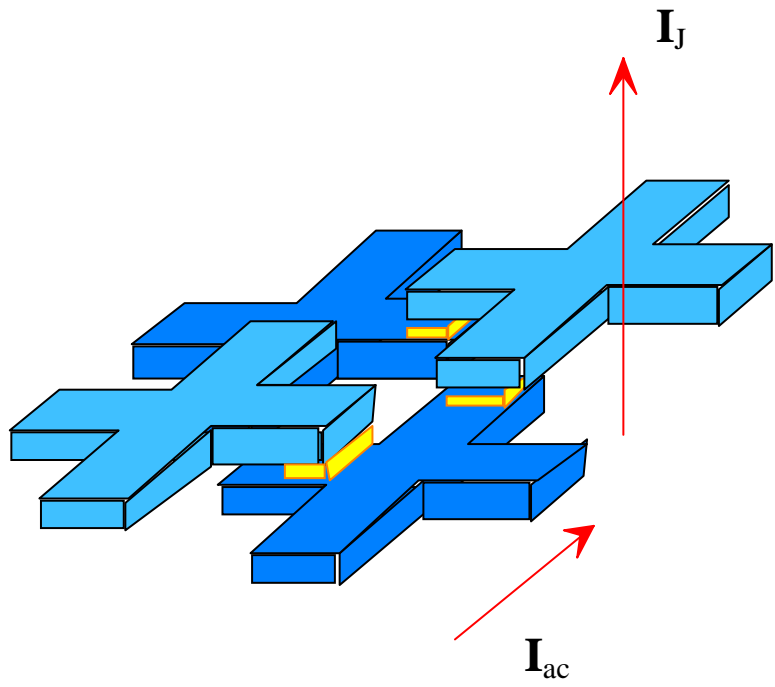

**Fig.2**

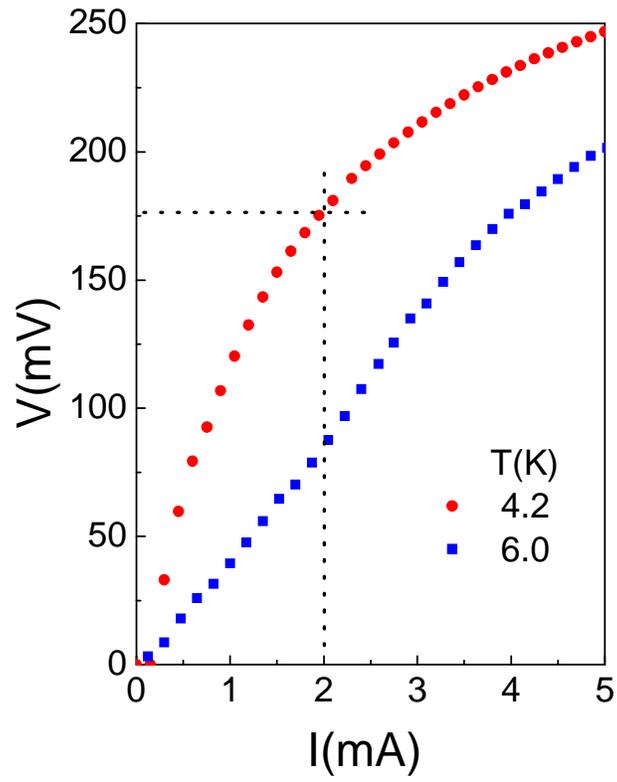

**Fig.3**

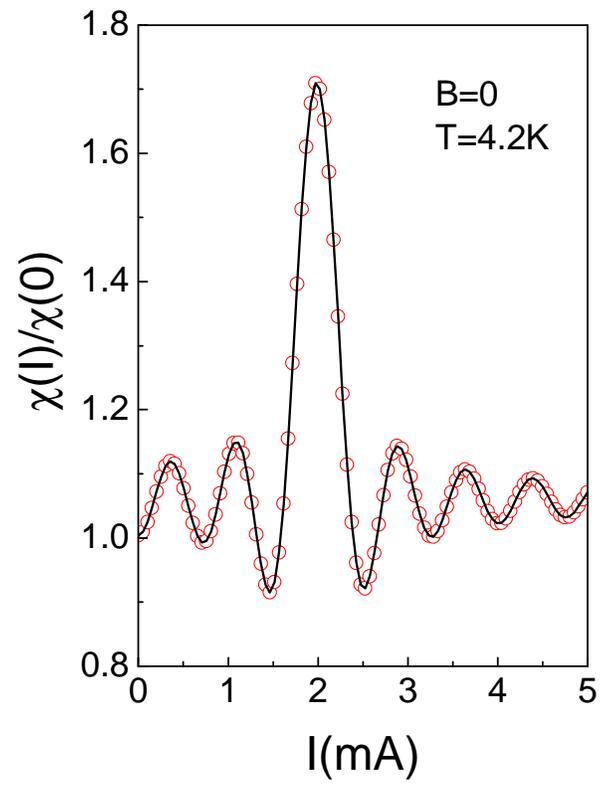

**Fig.4**